\documentclass[twocolumn]{aastex631}

\defcitealias{pearson16}{P16}

\shorttitle{HI in Dwarf-Dwarf Interactions}
\shortauthors{N. Luber et al.}
\graphicspath{{./}{figures/}}

\begin{document}

\title{Investigating the Baryon Cycle in Interacting Dwarfs with the Very Large Array and Pan-STARRS}

\correspondingauthor{Nicholas Luber}
\email{nicholas.m.luber@gmail.com}

\author{N. Luber}
\affiliation{Department of Physics and Astronomy, West Virginia University, P.O. Box 6315, Morgantown, WV 26506, USA}
\affiliation{Center for Gravitational Waves and Cosmology, West Virginia University, Chestnut Ridge Research Building, Morgantown, WV 26505}

\author[0000-0003-0256-5446]{Sarah Pearson}\thanks{Hubble Fellow}
\affiliation{Center for Cosmology and Particle Physics, Department of Physics, New York University, 726 Broadway, New York, NY 10003, USA}
\affiliation{Center for Computational Astrophysics, Flatiron Institute, 162 Fifth Avenue, New York, NY 10010, USA}

\author{Mary E. Putman}
\affiliation{Department of Astronomy, Columbia University, New York, NY 10027, USA}

\author{Gurtina Besla}
\affiliation{Steward Observatory, University of Arizona, 933 North Cherry Avenue, Tucson, AZ 85721, USA}

\author{Sabrina Stierwalt}
\affiliation{Physics Department, Occidental College, 1600 Campus Road, Los Angeles, CA 90041 USA}

\author{Joel P. Meyers}
\affiliation{Department of Astronomy, Columbia University, New York, NY 10027, USA}

\begin{abstract}

We present resolved HI synthesis maps from the Very Large Array (VLA) of three interacting dwarf systems: the NGC 3664 dwarf pair, the NGC 3264 dwarf pair, and the UGC 4638 dwarf triplet. All three dwarf systems are captured at various stages of interaction and span a range of environments. We detect clear hallmarks of tidal interactions through the presence of HI bridges, and diffuse HI extensions that surround the dwarfs. We overlay the HI data on Pan-STARRS r-band images and find further evidence of tidal interactions through coincident distorted HI and tidal stellar features in NGC 3264 and UGC 4638, and an unwound spiral arm pointing towards its smaller companion in NGC 3264. In UGC 4638, both the gas and diffuse stars are extended to similar radii east of the primary, which could indicate that the smaller dwarf in the system has already completed one pass through the primary. We additionally find that our three systems, and those from the Local Volume TiNy Titans survey, are not HI deficient and thus the interaction has not resulted in a loss of gas from the systems. A comparison with non-interacting dwarf galaxies shows that the interactions have a significant impact on the kinematics of the systems. Our new resolved HI kinematics, combined with detailed stellar and HI morphologies, provide crucial constraints for future dynamical modelling of hierarchical mergers and the baryon cycle at the low-mass scale.

\end{abstract}


\section{Introduction} \label{sec:intro}

\quad Dwarf galaxies are prevalent at all epochs in the Universe \citep{karachentsev04} and mergers between dwarfs are expected to occur more frequently than for massive galaxies in a given volume \citep{fakhouri10,deason14}. Recently, a wealth of studies have investigated the mutual interactions between dwarfs and how such interactions affect the baryon cycle of dwarf galaxies. Through observational work, we now know that dwarf interactions lead to a significant enhancement in star formation (SF) rates if the dwarfs in the pairs are separated by $<$50 kpc (\citealt{stierwalt15}), and that this SF is wide-spread and clumpy (\citealt{privon17}). We additionally know that dwarfs with high SF rates are more likely to host tidal features, indicating a recent merger, and that dwarfs with surrounding tidal features are systematically bluer (\citealt{fong20}). Theoretical work confirms the picture that mergers and flybys can enhance star formation in dwarf-dwarf galaxy encounters, e,g. \citet{martin21}, as well as interactions of any standard Hubble type galaxy, e.g. \citet{dimatteo07}.

The neutral gas content in dwarfs has also been studied extensively in several large-scale HI mapping surveys, such as in \citet{hunter12} (LITTLE THINGS), \citet{begum08} (FIGGS), and \citet{swaters02}. \citet{bradford15} showed that non-paired dwarfs retain their gas in the field ($f_{\rm gas} >$ 0.3\footnote{$f_{\rm gas} = \frac{M{\rm gas}}{M_{\rm gas} + M_{\rm star}}$}), despite the dwarfs' ongoing star formation and shallow potential wells. \citet{stierwalt15} additionally showed that if interacting dwarfs reside far from a massive galaxy they have similar atomic gas fractions to their non-paired counterparts. As such, dwarf interactions and SF alone do not seem to remove a large fraction of gas in dwarfs. Instead, large-scale environmental effects such as ram-pressure or tidal stripping from a nearby massive galaxy appear to be ultimately what removes gas from dwarfs. In particular, \citet{stierwalt15} found that dwarf pairs $<$ 200 kpc from a massive galaxy (M$_* > 10^{10}$ M$_\odot$) have systematically lower gas fractions when compared to dwarf pair counterparts that are farther than 200 kpc from any massive galaxy. This is supported by findings by \citet{geha12}, who showed that only dwarfs within 1.5 Mpc of a massive galaxy are quenched of their SF.

\quad We have further insight into the nature of dwarf mergers from the nearby Magellanic System (MS), which serves as the template for ongoing dwarf-dwarf interactions. A bridge of gas connects the two galaxies, and a leading and trailing gaseous stream span $>100^\circ$ of sky. Together, these features constitute an extended HI distribution that represents the ongoing interaction. The origin of the extended gaseous features, however, has remained ambiguous since their discovery \citep{mathewson74,putman98}, which is tied to the difficulty in modelling the dynamics of the 3-body Large Magellanic Cloud (LMC), Small Magellanic Cloud (SMC), and Milky Way (MW) system. Many models have invoked tidal or ram pressure forces from the MW halo to create the gaseous features \citep[e.g.,][]{gardinernoguchi96,connors06,mastropietro05,diazbekki11}, while others create the gas streams largely through the interaction of the Magellanic Clouds (MCs) themselves in a first in-fall scenario \citep[e.g.,][]{besla10,besla12,2018ApJ...857..101P}, and by additionally including a hot corona \citep{lucchini20}. 

\quad LMC-SMC pairs, as well as dwarf pairs and dwarf groups, are rare at $z=0$, both in the field and near MW type hosts, both observationally and theoretically \citep{boylan11,gonzales13,stierwalt17,besla18}. However, $10\%$ of dwarfs are cosmologically expected to have undergone a 1:10 mass ratio merger since $z=1$ \citep{deason14}. Since the interaction timescale of dwarfs can be long \citep[6-10 Gyr;][]{besla12,pearson18}, it is likely that such interactions are not cosmologically negligible to the baryon cycle of low mass galaxies. Given the significant amount of gas removed from the MCs \citep{for14, fox14, donghia16}, establishing the physical mechanism responsible for removing the large amount of mass is critical for our understanding of the baryon cycle of low mass galaxies and the role of minor mergers in feeding the circumgalactic medium (CGM) of galaxies like the MW.

\quad While recent observational and theoretical efforts have given insight into the frequency of dwarf interactions and the associated star formation, we still have a limited understanding of the distribution of gas surrounding dwarf pairs beyond the Magellanic System. \citet{pearson16} (hereafter \citetalias{pearson16}) studied the gas distributions of 10 interacting dwarf pairs using resolved synthesis HI maps (the Local Volume TiNy Titans). They found that the gas distributions of dwarf pairs were systematically more extended when compared to a control sample of non-paired dwarf irregulars with the same stellar extents (see \citetalias{pearson16}, figure 7). Hence, data strongly suggest a scenario in which tidal interactions serve to move gas to the outskirts of dwarf galaxies. Interestingly, this gas remains bound to the dwarf pairs unless the pairs are in the vicinity of a massive galaxy (see also \citealt{pearson18}). If gas is tidally pre-processed and extended due to dwarfs' mutual interaction, this can affect the efficiency at which gas is fed to more massive galaxies if the dwarfs are subsequently accreted. This was demonstrated by \citet{marasco16}, who used the EAGLE cosmological hydrodynamical simulations to show that dwarfs that had previously undergone a dwarf-dwarf encounter are more likely to be subject to efficient ram-pressure stripping, when in the vicinity of a more massive host galaxy. These results indicate that dwarf-dwarf interactions are an important part of the baryon cycle of low mass galaxies, which enables the parking of gas at large distances. This extended gas can serve as a gas supply channel to the dwarfs until accretion by a more massive host prevents this gas from being re-accreted by the dwarf pair. \citetalias{pearson16} consisted of a sample of only 10 pairs and used archival data that had varying sensitivity and resolution. Additional sensitive, high resolution observations are needed to confirm the results and expand the parameter space of the types of interactions.

\quad In this paper, we expand the sample of confirmed low-mass interacting galaxies that are mapped with high resolution HI data. We chose three systems that were candidate interacting dwarf systems, and mapped them in detail with Very Large Array (VLA) HI data. Our primary goals were to first confirm the interaction and then to characterize dwarf interactions in a range of environments.

\quad The NGC 3664 and NGC 3264 dwarf pairs were selected because they reside in an isolated environment and therefore provide insight to the mutual interaction between dwarfs irrespective of any nearby massive galaxy. These pairs have different pair separations than the pairs in \citetalias{pearson16}, and thus extend their sample to include pairs at different stages of interaction. The dwarf triplet UGC 4638 is located $<$200 kpc from a massive MW type spiral, and we selected this target to study the early stages of an interaction between a dwarf group and a more massive host galaxy. These data are a rare example of an HI map of an interacting dwarf triplet \citep[see also][]{chengalur13}. The high resolution resolved HI data provide unique insight into the baryon cycle and hierarchical structure formation at the low mass end \citep[see an example of an optical study of a dwarf group in][]{stierwalt17}.

\quad While we know of $>100$ dwarf galaxy pairs in the local universe ($z<0.02$) \citep[e.g.,][]{paudel18}, only 10s of dwarf pairs have high resolution resolved HI maps \citepalias[e.g.,][]{pearson16}. Adding three interacting dwarf systems to the existing sample of interacting pairs with detailed HI maps is therefore a significant increase. We search for evidence of a mutual interaction through the presence of HI bridges, which are hallmarks of interaction \citep[e.g.,][]{toomre72,combes78,hibbard95}. Additionally, we combine the HI data with the released Pan-STARRS data \citep{panstarrs} to reveal coincident tidal stellar and HI features, and investigate the gas content and detailed kinematics of the three systems. Our dwarf pairs thus contribute to the emerging observational and cosmological findings of how hierarchical interactions impact the baryon cycle of low mass galaxies.

\quad This paper is structured as follows: in Section \ref{sec:obs} we discuss the observing, calibration, and imaging strategies for the HI data. In Section \ref{sec:results}, we present the final HI moment maps for the three systems, as well as comment on the properties of these maps. In Section \ref{sec:disc}, we compare the HI morphology and kinematics of each system to a sample of isolated dwarf galaxies and discuss the HI and optical properties of the UGC 4638 triplet. Lastly, in Section \ref{sec:conc} we offer concluding remarks as well as motivation for future work that utilizes these data. 

\quad In this work, we assume a standard flat FLRW cosmology with H$_{0}$ = 70 km s$^{-1}$ Mpc$^{-1}$, $\Omega_{\Lambda}$ = 0.7 and $\Omega_{M}$ = 0.3.

\begin{deluxetable*}{cccccc}[t!h!]
\tablecaption{Summary of the Data}
\tablecolumns{6}
\tablenum{1}
\tablewidth{0pt}
\tablehead
{
\colhead{Target Field} &
\colhead{Array Configurations\tablenotemark{a,b}} &
\colhead{Time on Target} &
\colhead{R.M.S.\tablenotemark{c}} &
\colhead{Column Density Sensitivity\tablenotemark{d}} &
\colhead{Synthesized Beam\tablenotemark{e}} \\
\colhead{} & 
\colhead{} &
\colhead{hours} &
\colhead{$\mu$Jy beam$^{-1}$} &
\colhead{$\times$10$^{19}$ atoms cm$^{-2}$} &
\colhead{$\arcsec$}
}
\startdata
NGC 3664 & C + D & 4.00 & 315 & 0.78 & 36.0 $\times$ 33.0 \\
NGC 3264 &  C  & 2.75 & 300 & 1.13 & 31.0 $\times$ 25.0 \\
UGC 4638 & C + D & 3.50 & 300 & 2.84 & 20.0 $\times$ 15.5 \\
\enddata
\tablenotetext{a}{These observations were made during VLA semester 2017A, spanning from 2017-02-010 to 2017-08-28}
\tablenotetext{b}{Each observation had the identical correlator set up with a total bandwidth of 4MHz with 256 15.625 kHz channels.}
\tablenotetext{c}{Per 62.5 kHz emission free channel.}
\tablenotetext{d}{Per 62.5 kHz emission free channel, at the 2$\sigma$ level.}
\tablenotetext{e}{After imaging with robust weighting of 0.5 and smoothing for optimal sensitivity.}
\label{tab:obs}
\end{deluxetable*}

\begin{deluxetable*}{ccccc}[t!h!]
\tablecaption{Summary of System Properties}
\tablecolumns{5}
\tablenum{2}
\tablewidth{0pt}
\tablehead
{
\colhead{System} &
\colhead{Distance\tablenotemark{a}} &
\colhead{Pair Separation\tablenotemark{b}} &
\colhead{Distance to Host\tablenotemark{c}} &
\colhead{Primary Dwarf V-Band Magnitude\tablenotemark{d}} \\
\colhead{} & 
\colhead{Mpc} &
\colhead{kpc} &
\colhead{kpc} &
\colhead{mag}
}
\startdata
NGC 3664 & 19.7 & 35 & $>$ 1500 & -19.40 $\pm$ 0.41 \\
NGC 3264 & 13.4 & 12 & $\approx$ 800 & -19.34 $\pm$ 0.59 \\
UGC 4638 & 47.8 & $<$15& 180 & -19.16 $\pm$ 0.64 \\
\enddata
\tablenotetext{a}{The dwarf systems' luminosity distances are set by the systemic velocity of the dwarf systems as measured by the HI spectrum.}
\tablenotetext{b}{Projected physical separation between the dwarfs in the system calculated at their reported distance.}
\tablenotetext{c}{More massive host galaxy defined as a galaxy with $M_* > 10^{10}$M$_{\odot}$.}
\tablenotetext{d}{The absolute magnitude of the primary galaxy, defined as the most massive dwarf galaxy in the system, as reported by the NASA/IPAC Extragalactic Database (NED) when queried with the system ID.}
\label{tab:bkgprop}
\end{deluxetable*}

\section{Observations \& Data} \label{sec:obs}
\subsection{Targets}

\quad We observed the candidate interacting dwarf systems NGC 3664, NGC 3264, and UGC 4638 with the Karl G. Jansky Very Large Array (VLA) in the VLA-C and VLA-D configurations, as part of VLA project 17A-296. The VLA-C and VLA-D array have maximum baseline lengths of approximately 0.3 km and 1.0 km, respectively, resulting in sensitivity to emission on different spatial scales. NGC 3264, NGC 3664, and UGC 4638 are at different redshifts resulting in maximum sensitivity to different spatial scales. Peak sensitivity for our targets for C and D array are 0.9 and 3.0 kpc for NGC 3264, 1.3 and 4.4 kpc for NGC 3664, and 3.2 and 10.4 kpc for UGC 4638. As a result, combining these two configurations allows us to resolve the HI structure of the systems, while also detecting any extended HI morphological features. Due to scheduling, each target had a different number of observations, array configurations, and total target observation time, which affected the resolution and column density sensitivity. We have the best column density sensitivity for NGC 3664, allowing us to detect the most diffuse emission in that system. These specifics of the observations are recorded in Table \ref{tab:obs}. 

\quad The primary dwarf galaxy in our interacting systems span a range in total V-band magnitude of -19.16 to -19.40, comparable magnitudes to the LMC and M33 in the Local Group that have V-band magnitudes of -18.3 and -19.4, respectively\footnote[1]{The absolute magnitude of the primary galaxy, defined as the most massive dwarf galaxy in the system, as reported by the NASA/IPAC Extragalactic Database (NED) when queried with the system ID.}. The interacting dwarf system NGC 3664 was previously observed by \citet{wilcots04}. Our observations, have almost twice the spatial resolution and sensitivity. This increase in sensitivity allows us to further explore the HI bridge first tentatively detected in \citet{wilcots04}. Similarly, NGC 3264 was previously observed by \citet{wilcots96}, but this time our new observations offer a factor 3 greater spatial resolution and a factor 2 in sensitivity. This increased sensitivity allows us to potentially detect more diffuse HI to trace an interaction. No previous interferometric HI observations have been taken to understand the gas in the UGC 4638 system, which allows us to provide a first important look into the dynamics of a triple dwarf interaction.

\begin{deluxetable*}{cccccc}[t!h!]
\tablecaption{Neutral Hydrogen Properties}
\tablecolumns{6}
\tablenum{3}
\tablewidth{0pt}
\tablehead
{
\colhead{System} &
\colhead{Central Velocity\tablenotemark{a}} &
\colhead{W$_{20}$\tablenotemark{a}} &
\colhead{VLA HI Mass\tablenotemark{b}} &
\colhead{Single Dish HI Mass\tablenotemark{c}} &
\colhead{Mass Ratio of Members\tablenotemark{d}} \\
\colhead{} & 
\colhead{km s$^{-1}$} &
\colhead{km s$^{-1}$} &
\colhead{$\times$10$^{9}$M$_{\odot}$} &
\colhead{$\times$10$^{9}$M$_{\odot}$} &
\colhead{}
}
\startdata
NGC 3664 & 1376 & 159 & 1.65 $\pm$ 0.08 & 1.74 $\pm$ 0.01\tablenotemark{e} & 4:1 \\
NGC 3264 & 935 & 146 & 0.47 $\pm$ 0.04 & 0.63 $\pm$ 0.01\tablenotemark{f} & 90:1 \\
UGC 4638 & 3315 & 121 & 2.64 $\pm$ 0.15 & - & 4:2:1 \\
UGC 4640\tablenotemark{g} & 3312 & 337 & 5.90 $\pm$ 0.57 & 19 $\pm$ 0.01\tablenotemark{h} & $-$ \\
\enddata
\tablenotetext{a}{Velocity measurements taken from our VLA data and have an error of 13 km s$^{-1}$.}
\tablenotetext{b}{The combined VLA HI mass for the system measured over the entire region with emission (see Figure \ref{fig:HI_Maps}).}
\tablenotetext{c}{Single dish instruments do not have the required resolution disentangle the masses of UGC 4638 and UGC 4640.}
\tablenotetext{d}{An approximate mass ratio found by comparing the ratio of the HI mass within the optical component of each member galaxy.}
\tablenotetext{e}{The flux used in this calculation is from the ALFALFA survey \citep{2011AJ....142..170H}.}
\tablenotetext{f}{The flux used in this calculation is from the entry for NGC 3264 in HyperLEDA \citep{hyperleda}.}
\tablenotetext{g}{This is the Milky Way type, spiral galaxy which is a likely host galaxy to the UGC 4638 dwarf triplet system.}
\tablenotetext{h}{The combined single dish HI mass for UGC 4640 and UGC 4638. Note that we cannot derive a single dish mass for the triplet alone.}
\label{tab:HIprop}
\end{deluxetable*}

\subsection{Data Reduction}

\quad We carried out the data reduction using the Common Astronomy Software Application (CASA) with standard calibration methods \citep{casa}. We did the calibration in an iterative process where we calculated an initial calibration, flagged Radio Frequency Interference (RFI) from the data, and then recalculated the calibration on the RFI-free data. We then flagged the target fields by hand with the automated flagging agent, \textit{RGLAG}, which is an internal flagging agent to CASA that flags out statistical outliers.

\quad After we completed the calibration, we set the weights for each dataset to be integration based in order to properly combine the data for imaging. We then frequency averaged to achieve a spectral resolution of 62.5 kHz, which corresponds to a channel width of 13 km s$^{-1}$ at $z = 0$. Next, we imaged the data with a robust weighting of 0.5 in order to create a cube of optimal sensitivity and resolution. Finally, we identified the emission-free channels and used them to fit a first order polynomial using the CASA task \textit{IMCONTSUB} to produce a continuum free cube. The properties of the three cubes are summarized in Table \ref{tab:obs}.

\subsection{Final Moment Maps}

\quad In order to best define the region containing HI emission, we performed a thorough inspection of emission persistence across different spatial and spectral smoothing combinations. We identified the HI emission in cubes of different resolutions, and determined a final mask for the detection by comparing the different resolution cubes and extrapolating them to the full resolution cube. We created the final HI maps by using the cube of optimal resolution for the HI emission and integrating over all pixels within the mask with values greater than $\pm$2$\sigma$. This strategy for moment map creation is implemented to ensure that the lowest contour is real emission and remains as clean as possible from RFI. We calculate HI masses by adding the flux in a region that is two pixels more extended than the lowest column density contour in the final moment map in every channel for which emission is present. This method of mass calculation is sensitive to the mass contribution for even the lowest level emission, allowing for the most accurate mass measurement.

\quad The dwarf systems' luminosity distances were set by the HI systemic velocity of the dwarf systems. We estimated HI mass errors by taking the flux of regions in random locations of identical size to the one used for the mass calculation and find the standard deviation of the distribution of these empty field masses. 

\section{Results}\label{sec:results}

\quad In this section, we present the final moment maps and PAN-STARRS optical images for each of the interacting systems, as well as a qualitative description of their HI emission. We summarize the properties of our targets in Tables \ref{tab:bkgprop} and \ref{tab:HIprop}.

\begin{figure*}
\begin{center}
\includegraphics[scale=0.45]{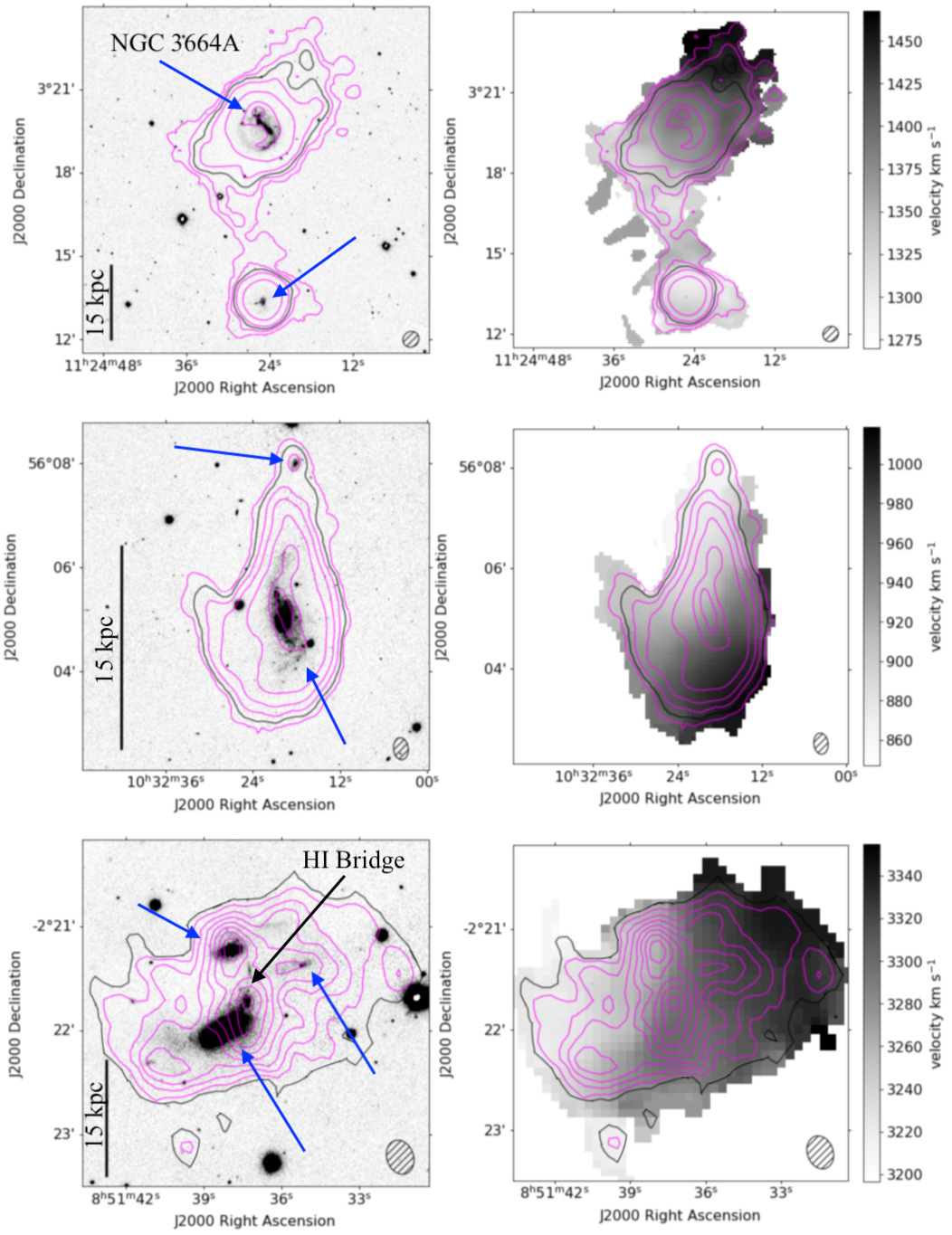}
\caption{The neutral hydrogen morphology (left) and kinematics (right) for the NGC 3664 (top), NGC 3264 (middle) and UGC 4638 (bottom) dwarf systems, overlaid on Pan-STARRS r-band images \citep{panstarrs}. The magenta lines correspond to column densities of: \textit{Top:} 0.13, 0.40, 1.20, 3.60, 10.8, 15.00 x 10$^{20}$ atoms cm$^{-2}$. \textit{Middle:} 0.35, 1.3, 2.5, 5.0, 10.0, 11.9 x 10$^{20}$ atoms cm$^{-2}$. \textit{Bottom:} 1.5,3,4.5,6,7.5,9,10.5,12 x 10$^{20}$ atoms cm$^{-2}$. The black contour in all plots corresponds to a column density of 7 x 10$^{19}$ atoms cm$^{-2}$. The blue arrows point to the location of the dwarf members in each system. All images are oriented with north pointing up, and east pointing left, and the beams are the hatched ellipses appearing in the south west corners.}
\label{fig:HI_Maps}
\end{center}
\end{figure*}

\vspace{10mm}

\subsection{The NGC 3664 system}

\quad In the top row of Figure \ref{fig:HI_Maps}, we present the HI morphology (left) as well as the HI kinematic information (right) for NGC 3664 overlaid on a Pan-STARRS r-band image. The NGC 3664 system consists of two dwarf galaxies separated by roughly 38 kpc in projection, and is isolated ($>$1.5 Mpc) from any galaxy with a stellar mass of more than M$_* >$ 10$^{10}$ M$_{\odot}$. Our HI map (Fig. \ref{fig:HI_Maps}, left) shows that the barred Magellanic-type spiral and the small irregular companion (see blue arrows) are connected by a low column density HI bridge (see purple contours). \citet{wilcots04} previously presented HI observations of the NGC 3664 system, and their data were highly suggestive of a relationship between the two galaxies, but the HI column density was not sensitive enough to connect the objects. 

\quad The connecting bridge between the two dwarf galaxies confirms they are interacting. HI bridges are a common consequence of galactic tidal interactions \citep[e.g.,][]{hibbard95, privon13}. We note that the NGC 3664 bridge has a relatively low column density, $\approx$ 2 $\times$ 10$^{19}$ atoms cm$^{-2}$, compared to the interacting dwarf pairs in \citetalias{pearson16}, in which all bridges have column densities greater than 8 $\times$ 10$^{19}$ atoms cm$^{-2}$. However, any bridges that do exist at the low level of NGC 3664's bridge would not have been detected in the \citetalias{pearson16} sample due to the poorer sensitivity. The HI bridge is likely too diffuse for star-formation to occur, as in the case of the Magellanic Bridge, and in fact in Figure \ref{fig:UVimages} (top), we see no evidence of UV emission in the location of the HI bridge. The UV emission in both the primary and secondary galaxy closely follows the optical emission in the Pan-STARRS images. This indicates that while there is star-formation in the optical extent of the galaxies, there is no detectable star-formation in the diffuse HI bridge connecting the systems. This is consistent with the fact that the separation between these two galaxies is at least twice that of the Magellanic Clouds. The lack of a tidal stellar bridge feature is likely due to the fact that HI is typically more extended than stars in dwarfs \citep[e.g.,][]{swaters02} and therefore gets affected tidally first. 

\quad The total HI mass within the HI bridge, defined as all gas between declinations of 3$^{\circ}16\arcmin$ - 3$^{\circ}17\arcmin$, which is a projected length of 5.7 kpc, is M$_{\rm HI}\approx$ 3 $\times$ 10$^{7}$ M$_{\odot}$, and has a peak HI column density of 4 $\times$ 10$^{19}$ atoms cm$^{-2}$. For comparison, the total HI mass within the region shown in the upper left panel of Figure \ref{fig:HI_Maps}, is M$_{\rm HI} = $ 1.65 $\pm$ 0.08 $\times$ 10$^{9}$ M$_{\odot}$ (see Table \ref{tab:HIprop}). Thus, the bridge HI mass only contributes a few percent to the total HI mass of the system. The total HI mass within the primary (NGC 3664A), which is the most massive galaxy in NGC 3664, defined as all the HI in the region at a declination $> 3^{\circ}$16$\arcmin$, is M$_{\rm HI} = $1.54 $\times$ 10$^{9}$ M$_{\odot}$.

\quad The HI distribution and the kinematics around NGC 3664A offers further evidence of the interaction between these two dwarfs: the HI is significantly more elongated along the velocity major axis than the minor axis (Figure \ref{fig:HI_Maps}, top). This is unusual as the kinematic major axis is not aligned with the optical major axis. However, the optical morphology is disturbed to the southeast, which is in the direction of the HI kinematic axis. The kinematic major axis is not pointing towards the companion galaxy, which lies due south. This could be a consequence of the specific orbital properties of the encounter or due to an offset between the orbital plane and the inclination of the smaller dwarf (see e.g., \citealt{pearson18}). 

\quad The Pan-STARRS r-band image illustrates that the NGC 3664 system could be an example of a generic collision scenario leading to an one-armed, off-centered, barred galaxy \citep{athanassoula96, berentzen03,bekki09, besla12, besla16, pardy16}. Additionally, the primary, more massive galaxy, NGC 3664A, does not show evidence for differential rotation (Figure \ref{fig:HI_Maps}, top right, and Figure \ref{fig:pv_slice}, left panel). The velocity gradient across the kinematic major axis is increasing, rather than reaching a peak on either side of the galaxy and then flattening out. 

\subsection{The NGC 3264 system}

\quad In the middle row of Figure \ref{fig:HI_Maps}, we present the HI morphology (left), as well as the HI kinematic information (right) for the NGC 3264 system overlaid on Pan-STARRS r-band data. This system is an example of a more massive dwarf galaxy accreting a smaller dwarf companion (see blue arrows), with an approximate mass ratio of 90-to-1 (see Table \ref{tab:HIprop}. The projected separation between the two dwarfs is only $\approx$ 12 kpc and the system is relatively isolated, approximately 800 kpc from the nearest massive galaxy. 

\quad We measure a total HI mass for the entire NGC 3264 system of 0.47 $\pm$ 0.04 $\times$10$^{9}$ M$_{\odot}$ (see Table \ref{tab:HIprop}) across the region showed in Figure \ref{fig:HI_Maps}. The single dish HI measurement found from \citet{hyperleda} for this system is higher at 0.63 $\pm$ 0.01 $\times$10$^{9}$ M$_{\odot}$ indicating there is additional diffuse gas that is not recovered here. The most striking feature of this system, is the elongation along the east side of the kinematic minor axis (at RA $\approx$ 10h32m28s). This may be an induced tidal feature as a result of a specific orbital configuration of the smaller secondary dwarf around the larger primary dwarf. 

\begin{figure}
\begin{center}
\includegraphics[scale=0.25]{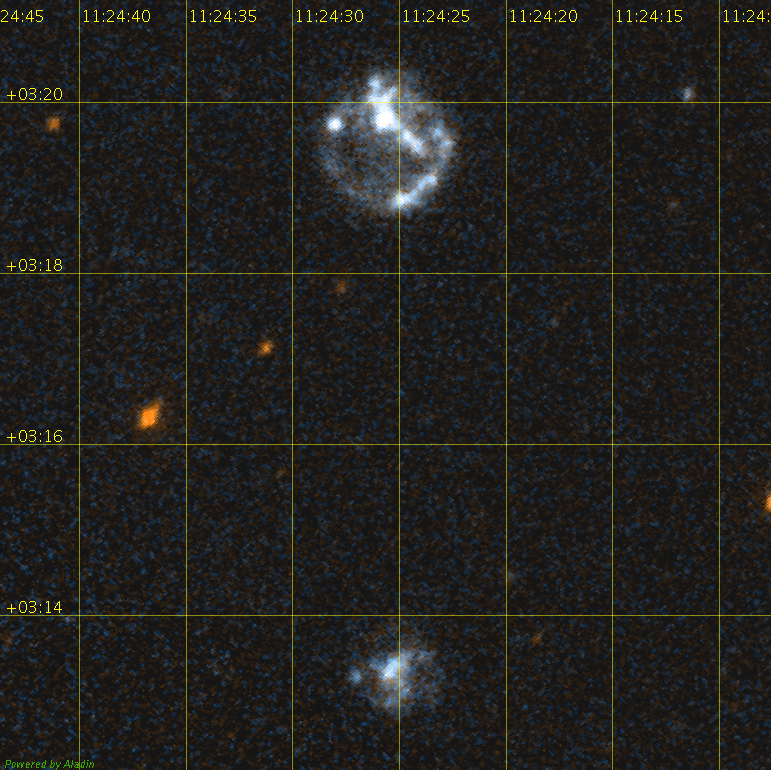}
\includegraphics[scale=0.25]{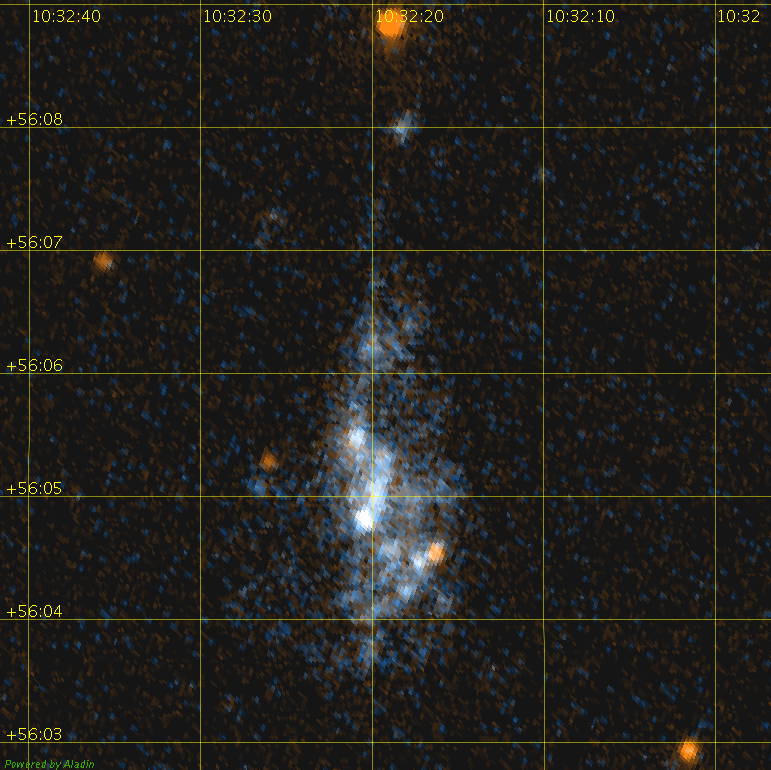}
\includegraphics[scale=0.25]{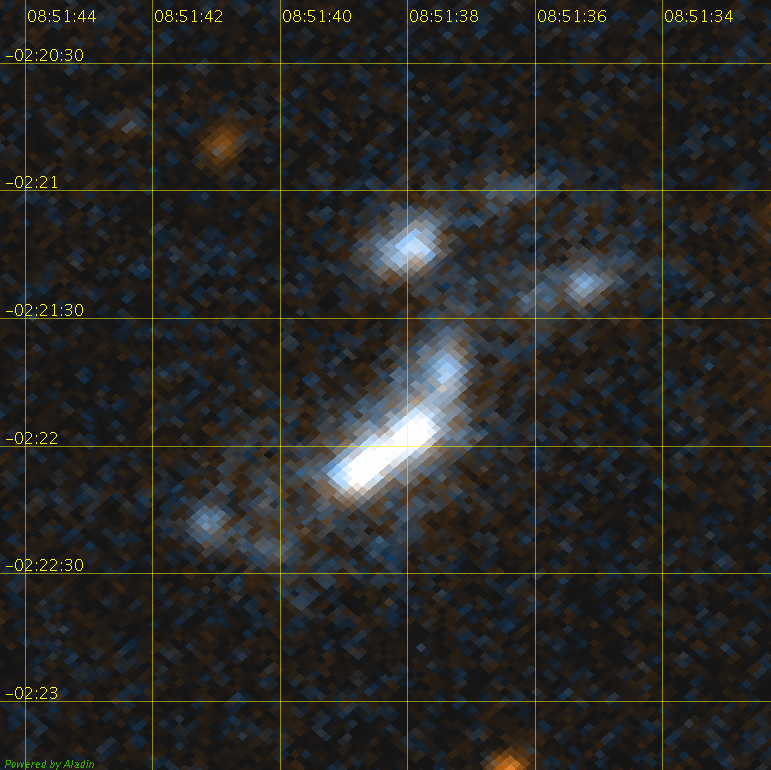}
\caption{Ultraviolet images for \textit{Top:} NGC 3664, \textit{Middle:} NGC 3264, and \textit{Bottom:} UGC 4638, with a field of view of 2.58$\arcmin$, centered on XX. The images are taken from the GALEX Allsky Imaging Survey (AIS) \citep{galex}. In each image, the transfer function is the native color map produces by the GALEX AIS team, as outputted by the Aladin Sky Atlas \citep{aladin1,aladin2}.}
\label{fig:UVimages}
\end{center}
\end{figure}

\quad In \citet{wilcots96}, the authors showed the same HI extension along the north-south direction, but due to resolution limits, they were unable to resolve the smaller companion within the dwarf system. Our observations have the resolution and sensitivity to show a clear secondary peak in HI emission coincident with the smaller companion in the r-band Pan-STARRS image. Our observations show that these galaxies are connected by the lowest column density contour with a smooth velocity gradient (Figure \ref{fig:HI_Maps}, middle right). This is clearly an interacting system with an HI mass ratio of 90 to 1. In the high resolution Pan-STARRS images, we see that the northern spiral arm of the primary galaxy, NGC 3264, appears to be at a different pitch angle than that of the arm in the south, which is an indication of a spiral galaxy being unwound \citep{bellhouse21}. Additionally, we see this feature more pronounced in the GALEX image (Figure \ref{fig:UVimages} middle) with the UV emission even more extended than the optical emission in this unwinding arm. The fact that there is more UV emission than optical emission in this unwinding arm indicates that the stellar population is younger, and that it is the location of active star-formation. This galaxy unwinding in the direction of the smaller interacting companion is consistent with a tidal interaction between the two dwarfs \citep[e.g.,][]{toomre72}. The recent star-formation in this unwinding arm is an indication that the tidal interaction between the two galaxies is responsible for the conversion of the HI envelope connecting the systems into new stars.

\begin{figure*}
\includegraphics[width=\textwidth]{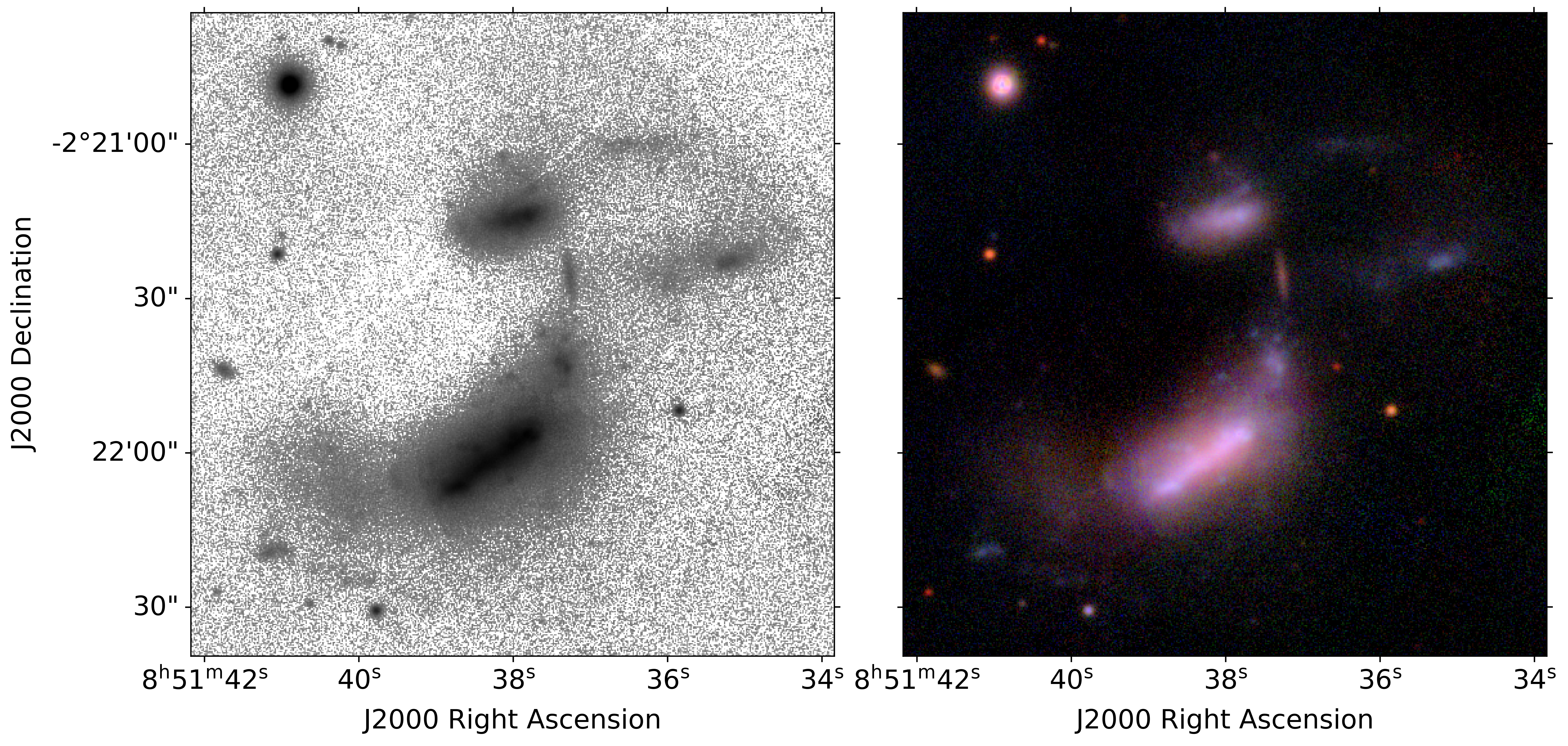}
\caption{\textit{Left:} A zoom in on the Pan-STARRS $r$-band image of the UGC 4638 dwarf triplet. The grey-scale is logarithmic, the window has a physical size of 28.3 x 23.8 kpc. We clearly see extended diffuse starlight coming off of the primary galaxy (lower left corner), which is similar to the diffuse extended starlight from the NGC 4490 dwarf in \citet{pearson18}. We also see clumpy star formation north of the primary, as well as diffuse starlight extending from the two smaller dwarfs in the central west and north west (upper right) part of the image. \textit{Right:} Same as left, but showing a Pan-STARRS $i, r, g$-band composite image of the UGC 4638 triplet, with the color scale representing from red to blue, corresponding to the intensity of the bands $i, r, g$, in that order, with the transfer function weighted more toward the $g$-band, such that fine color distinctions can be made. This image reveals that the diffuse stellar extensions follow the overall color distribution of the galaxies and is therefore likely associated with the triplet. The thin, dense stellar feature between the three dwarfs is redder and appears to have a central bulge. It is therefore likely a background galaxy.}
\label{fig:UGC4638_zoom}
\end{figure*}

\subsection{The UGC 4638 system}

\quad In the bottom row of Figure \ref{fig:HI_Maps}, we present the first maps of the total HI morphology (left), as well as the HI kinematic information (right) for UGC 4638: a system of three dwarfs (see blue arrows) interacting near the more massive Milky Way type spiral galaxy, UGC 4640. The HI contours are overlaid on a Pan-STARRS r-band image. The extent of the HI envelope for the three interacting dwarfs is 40 kpc in projection, and the three dwarfs reside 180 kpc in projection from the more massive UGC 4640 spiral galaxy. The total HI mass of the triplet, measured over the region shown in Figure \ref{fig:HI_Maps}, is 2.64 $\pm$ 0.15 $\times$10$^{9}$ M$_{\odot}$ (see Table \ref{tab:HIprop}). 

\quad There are some faint extended optical features between the two smaller mass members (in the north) which are likely results of an interaction. In Figure \ref{fig:UVimages} (bottom), we see that the UV emission is distributed through all areas where we see optical emission, indicating that the entire system is undergoing current star formation. The diffuse UV emission is also coincident with the diffuse HI emission, suggesting that young stars are being formed in the diffuse HI enveloping the systems. This indicates that the interaction is causing star-formation outside of the main bodies of the constituent dwarf galaxies.
 
\quad The diffuse starlight and HI in the southeast extending from the primary dwarf appears similar to the extended HI and stars from the primary dwarf in the NGC 4490/85 pair (see \citealt{pearson18}, fig. 1). Through dynamical modeling, \citet{pearson18} found that the smaller dwarf, NGC 4485, had passed through the disk of NGC 4490 and induced a one-armed spiral, which we view edge-on. We might be seeing a similar scenario for the triplet system, which is supported by the presence of the HI bridge. Another scenario could be that the coincident extension of the HI and the diffuse starlight east of the primary in the triplet indicates a late-stage interaction. In an early-stage interaction, only the HI would be disturbed, because the stars are typically less extended than the gas (\citealt{swaters02}) and have a stronger anchoring force. In a later-stage interaction, however, the stars could have had time to be pulled out by the tidal forces, or have had time to form in the gas stripped from the dwarf system. However, dynamical models are necessary to make any definitive conclusions about the dynamical state of the triplet system. 

\begin{figure}
\includegraphics[width=\columnwidth]{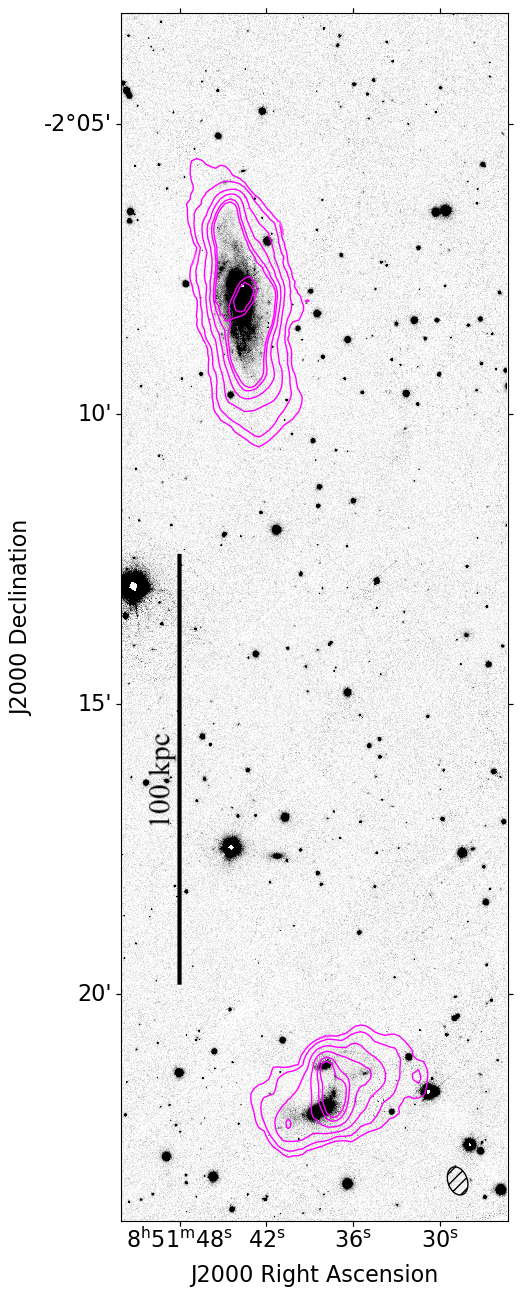}
\caption{The HI morphology overlaid on a Pan-STARRS r-band image of the UGC 4638 dwarf triplet (bottom) and the nearby UGC 4640 spiral galaxy (top). The contours correspond to HI column density levels of 0.93, 1.85, 3.70, 5.55, 7.40, 8.79 x 10$^{20}$ atoms cm$^{-2}$.}
\label{fig:UGC4640}
\end{figure}

\quad Our original VLA pointing was centered in-between the UGC 4638 system and the massive galaxy UGC 4640, allowing for both galaxies to reside within the primary beam of the VLA pointing, and accurate mapping and mass measurements to be made. In Figure \ref{fig:UGC4640}, we again show the HI distribution for the UGC 4638 triple (south) but also the nearby massive galaxy UGC 4640 (north). The total HI mass of UGC 4640 is 5.9 $\pm$ 0.57 $\times$10$^{9}$ M$_{\odot}$ (see Table \ref{tab:HIprop}). The triplet's systemic velocity is less than 10 km s$^{-1}$ offset from the UGC 4640 galaxy's systemic velocity (see Table \ref{tab:HIprop}), and is located only 180 kpc away in projection. These two facts imply that the UGC 4638 dwarf triplet is likely dynamically associated with the larger spiral, similar to the Milky Way and Magellanic Clouds (see also \citealt{paudel17,paudel18,paudel20}). Note, however, that UGC 4638 resides three times farther from UGC 4640 than the MCs reside from the MW.

\quad We took follow-up observations for the UGC 4638 triplet and UGC 4640 spiral system using the Greenbank Telescope (GBT). We mapped a 45\arcmin~x 45\arcmin~region around the system and calibrated and imaged the data using standard packages in GBTIDL. No diffuse HI could be explicitly disentangled from the gas from the two bright HI sources due to the large beam of the GBT. This caused the sources to be coincident within the same lowest column density contour. We are able to calculate a total HI mass for the UGC 4638/4640 system of 1.9$\times 10^{10}$ M$_{\odot}$ from the GBT data. This is approximately double of the sum of the HI masses measured from the VLA data. This result suggests that a large proportion of the total neutral gas of the UGC 4638/40 system resides outside of the HI envelopes detected by the VLA. 

\section{Discussion}\label{sec:disc}

\begin{figure*}
\begin{center}
\includegraphics[width=\textwidth]{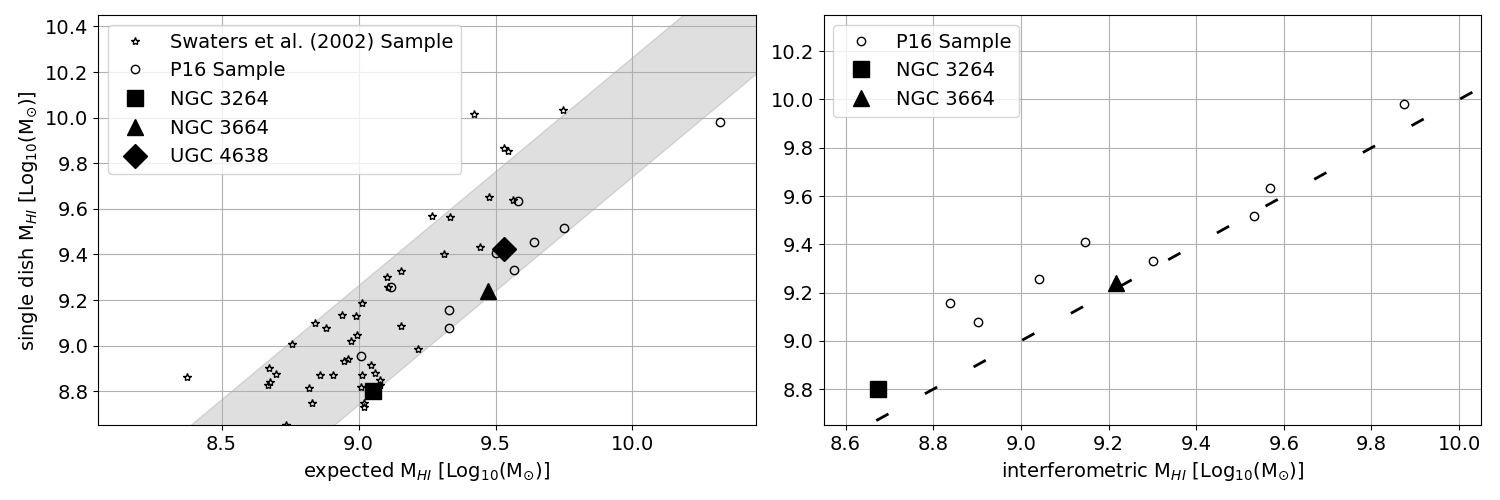}
\caption{\textit{Left:} The measured HI mass as a function of expected HI mass for the 10 systems in the \citetalias{pearson16} sample (black outlined circles), the \citet{swaters02} sample (black outlined stars), and the three dwarf systems NGC 3264, the square, NGC 3664, the triangle, and UGC 4638, the diamond. For all values except for UGC 4638, the HI measurements are from single dish telescopes. The expected value is derived through the HI mass/B-band optical radius relationship for dwarf galaxies in \citet{swaters02}, using Equation \ref{eq:hi_opt} and the values from Section \ref{sec:disc}. The gray shaded region illustrates the $\pm$1$\sigma$ error on the fit calculated from the 73 isolated dwarf galaxies with HI masses in the range 10$^{7.5-10.0}$ M$_{\odot}$ in \citet{swaters02}. Note that this figure only shows those galaxies with an observed HI mass greater than 10$^{8.7}$ M$_{\odot}$. The dwarf systems presented in this paper and the dwarf pairs from \citetalias{pearson16} do not appear significantly HI deficient when compared to the \citet{swaters02} sample of isolated dwarfs. \textit{Right:} The relationship between measured single dish HI masses and interferometric HI masses for the same systems. The interferometric measurements for the \citetalias{pearson16} sample come from their Table A2 and the single dish measurements are compiled from the literature \citep{putman03,hipass,koopmann08,matthews08,bottinelli78,richter94,shostak80}. The values for NGC 3264 and NGC 3664 are those reported in Table \ref{tab:HIprop}. The single dish masses are generally larger than the interferometric masses, which means there is significant extended gas in these systems.}
\label{fig:Pred_Mass}
\end{center}
\end{figure*}

\begin{figure*}
\begin{center}
\includegraphics[width=\textwidth]{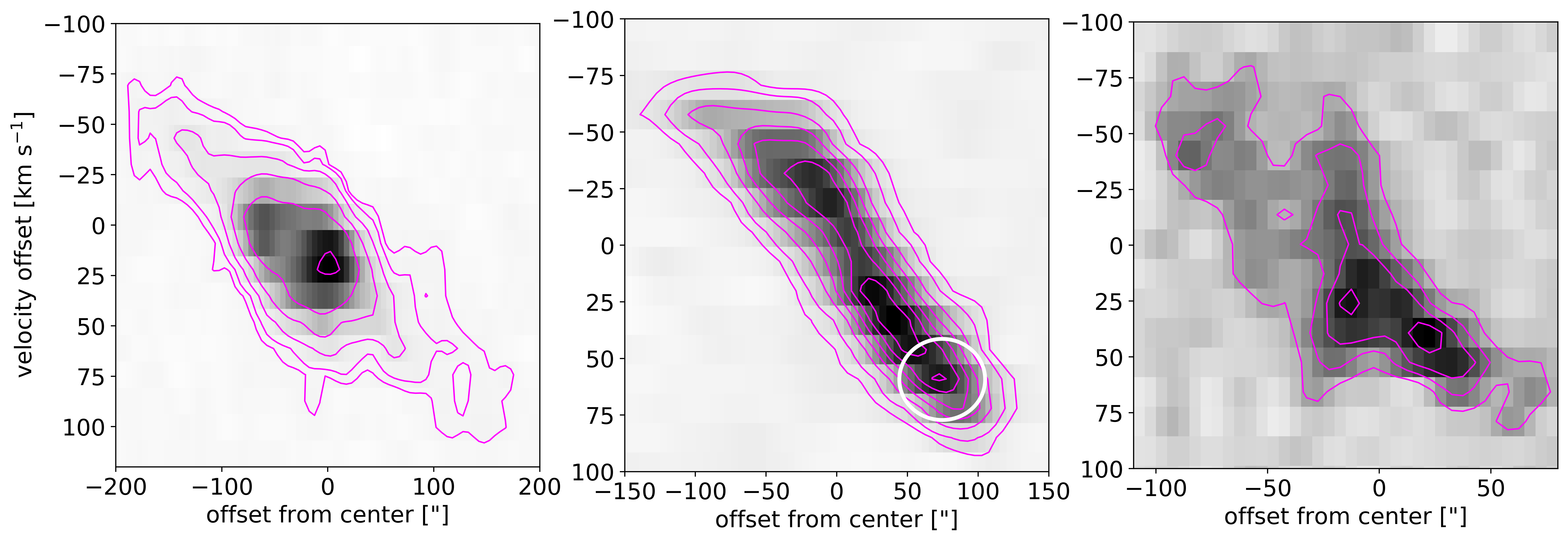}
\caption{Position-Velocity slices taken along the HI morphological and kinematic major axis for \textit{Left:} NGC 3664A, the barred-Magellanic primary galaxy in the NGC 3664 system, \textit{Middle:} the entire NGC 3264 system, and \textit{Right:} the entire UGC 4638 system. In all cases the contours are adjusted to best trace out the extended emission. The position-velocity slices are taken over the position angles, where the position angle is measured clockwise, going North-South, with 0$^{\circ}$ being North, 48$^{\circ}$, 5$^{\circ}$, and 62$^{\circ}$, for NGC 3664A, the NGC 3264 system, and the UGC 4638 system, respectively. The velocity offset axis is centered at systemic velocity for each system. In each case, the position-velocity slices do not show the characteristic flattening out at high offset from positional center. The white circle in the slice for NGC 3264 (middle) is the approximate location of the secondary galaxy, and the gas outside of this circle is all from the primary dwarf. The dense concentration of gas in the right panel represents the positions of the galaxies in the UGC 4638 system.}
\label{fig:pv_slice}
\end{center}
\end{figure*}

\quad Here we discuss several of the science implications that can be extracted from our detailed maps of the gas distribution for the three interacting dwarf systems. In Section \ref{sec:hicontent}, we investigate the total measured HI masses of our individual dwarf systems, and compare the values to what we would expect given the dwarfs' optical extents. In this analysis, we include a comparison to the \citetalias{pearson16} dwarf pair sample. In Section \ref{sec:n3664}, we analyze the detailed kinematic maps of the NGC 3664 and the NGC 3264 systems, and compare their velocity profiles to expectations from empirical relations. We investigate the kinematics and potential interaction of the UGC 4638 triplet with its more massive host, UGC 4640, in Section \ref{sec:triplet}.

\subsection{Investigating the Total to Expected HI Content of the Dwarf Systems}\label{sec:hicontent}

\quad The gas content of interacting dwarf galaxies should be dynamically affected by the interaction. To investigate whether this is true for the dwarfs presented in this work, we compare the HI content vs. diameter relationship of our dwarf systems to what is expected from isolated dwarfs. To determine what the expected HI mass is for isolated dwarf galaxies, we use the empirical relationship from \citet{swaters02}:

\begin{equation}
{\rm log}_{10}({\rm M}_{\rm HI}) = B + A \times {\rm log}_{10}(\rm D)
\label{eq:hi_opt}
\end{equation}

where D is a dwarf's HI diameter, taken to a column density of 1 M$_{\odot}$ pc$^{-2}$, and M$_{\rm HI}$ is the total expected HI mass based on this diameter. The fit to the sample of 73 isolated dwarf galaxies with HI masses in the range 10$^{7.5-10.0}$ M$_{\odot}$ in \citet{swaters02}, gives values of $A = 1.43$ and $B = 7.25$ for the free parameters in Equation \ref{eq:hi_opt}. The masses for the isolated dwarf galaxies in \citet{swaters02} are derived using single dish measurements and are therefore representative of the total HI mass in the system. 

\quad For each dwarf in our three dwarf systems, we measure the HI diameter out to 1 M$_{\odot}$ pc$^{-2}$ using our interferometric maps. We then compute the expected HI mass for each dwarf galaxy using Equation \ref{eq:hi_opt}. Subsequently, for each particular dwarf system, we combine the expected HI mass from each dwarf galaxy into a total expected HI mass for the entire system. Thus, for NGC 3664 and NGC 3264, we combine the expected HI mass for the two dwarfs in each system, and for UGC 4638, we combine the expected HI mass for the three dwarfs. To compare our results to the findings in \citetalias{pearson16}, we additionally used the HI extent, at a column density of 1.04 M$_{\odot}$ pc$^{-2}$\footnote[1]{We are unable to use exactly 1 M$_{\odot}$ pc$^{-2}$ but the difference in radius for this slight difference in column density is negligible.} radii, from their Table A.2, and followed the procedure above to obtain the expected HI mass for the 10 dwarf systems in the their sample. 

\quad In Figure \ref{fig:Pred_Mass} (left), on the x-axis we present the HI mass from Equation \ref{eq:hi_opt}, and show the relation with the observed single dish HI mass on the y-axis for the \citetalias{pearson16} sample, the NGC 3264 system, and the NGC 3664 system (see Table \ref{tab:HIprop}). For UGC 4638, we cannot derive a single dish mass (see Table \ref{tab:HIprop} for a combined UGC 4638 and 4640 systems), thus for UGC 4638 we plot the VLA HI mass. We plot our three interacting dwarfs systems as a square (NGC 3264), triangle (NGC 3664), and diamond (UGC 4638), and we present the \citetalias{pearson16} sample as black outlined circles. Each point corresponds to an entire system (i.e. including the primary, secondary and, for UGC 4638, the tertiary dwarf). 

\quad The shaded grey region in Figure \ref{fig:Pred_Mass} (left) shows 1$\sigma$, as measured by calculating the residual r.m.s. about the fit, above and below the case in which the expected HI mass from Equation \ref{eq:hi_opt} is equivalent to the observed HI mass. The observed combined HI masses for the 13 dwarf systems are all within 2$\sigma$ of the expected HI mass, and do not appear deficient in HI when compared to the \citet{swaters02} control sample. This is consistent with findings of \citet{stierwalt15}, who showed that only dwarf pairs residing $<200$ kpc from a massive host galaxy can be HI deficient when compared to isolated, non-paired dwarfs. While the UGC 4638 system and two of the \citetalias{pearson16} pairs (see their Table 4) reside $<$200 kpc from a massive galaxy, based on Figure \ref{fig:Pred_Mass} (left) these systems do not yet appear to have lost a substantial fraction of their gas. Interestingly, when the constituent galaxies for NGC 3264 and NGC 3664 are considered individually, it is the smaller companion galaxies that are more deficient in their HI reservoir. In each case the secondary dwarf galaxies are deficient in mass by approximately a factor 3, while the primary galaxies are only slightly deficient. 

\quad In Figure \ref{fig:Pred_Mass} (right), we show the relationship between measured single dish HI masses and interferometric HI masses for the same systems (note that we omit UGC 4638 as we could not obtain a single dish HI mass for this system). The interferometric measurements for the \citetalias{pearson16} sample come from their Table A2. We see that the single dish values are generally higher than the interferometric HI masses, which means that gas resides in a diffuse envelope around the interacting system. If we had instead used HI measurements from interferometers on the y-axis of Figure \ref{fig:Pred_Mass} (left), the values from \citetalias{pearson16} and for our systems presented here would be systematically lower. The fact that the single dish HI mass measurements capture this gas, confirms the findings of \citetalias{pearson16}, who argued that gas is parked at large distances through dwarf interactions, but that gas is not fully removed unless the pairs are in the vicinity of a massive host. Using HI measurements from interferometers, can thus lead to an underestimate of the HI mass, as interferometers are not sensitive to large-scale emission. Our VLA HI mass for NGC 3264 is approximately 30\% lower than the single dish value reported in Table \ref{tab:HIprop}. Our HI mass measurement for NGC 3664 is in agreement with the single dish measurement shown in Table \ref{tab:HIprop}, with only a 5\% discrepancy in mass. From our GBT measurements, we know that the UGC 4638/40 system is more massive than our VLA measurements imply, but it is not possible to disentangle the triplet from the spiral. For UGC 4638 in Figure \ref{fig:Pred_Mass} (left) we plotted its y-axis observed mass from interferometric data. On average, the interferometric measurements for the \citetalias{pearson16} sample are 26\% lower than the single-dish measurements. From our GBT observations including UGC 4640, we know the true mass is likely higher. This suggests that the point should be higher, making it among the most gas rich in our sample and that in \citetalias{pearson16}. This is perhaps a result of the unique triple nature UGC 4638 system, or perhaps indication of an early-stage interaction with the massive galaxy UGC 4640.

\subsection{The Peculiar Velocity Profiles of NGC 3664 \& NGC 3264}
\label{sec:n3664}

\quad In \citet{karachentsev17}, the authors use the kinematics, luminosities, and gas masses of dwarf galaxies to make a suite of different Tully-Fisher relationships. In this section, we investigate how our sample of dwarf-dwarf systems compares to the expected values in \citet{karachentsev17}, and what differences in the expected and measured values can imply about the nature of the dwarf-dwarf systems. 

\quad In the primary galaxy in NGC 3664, henceforth NGC 3664A, we measured a total velocity width of W$_{50}$ = 66 km s$^{-1}$ at 50\% of its maximum flux from our VLA data. Assuming an inclination angle of 57.2$^{\circ}$\footnote[1]{Measurement from the entry for NGC 3664 in Hyperleda \citep{hyperleda}.}, the total corrected velocity width is W$_{50}^{i}$ = 78.5 km s$^{-1}$. The total HI mass within just NGC 3664A, defined as all the HI in the region at declination $>3^{\circ}$16$\arcmin$, is 1.54 $\times$ 10$^{9}$ M$_{\odot}$. Using the \citet{karachentsev17} relation for the gaseous Tully-Fisher relationship for dwarfs, which relates HI mass and W$_{50}^{i}$ (see \citet{karachentsev17} Figure 9 for the relationship and accompanying data), we find that NGC 3664A's total HI content derived from its W$_{50}^{i}$, should be 1.16$\times$10$^{8}$ M$_{\odot}$, which is an order of magnitude lower than the observed mass. Equivalently, the system is rotating 100 km s$^{-1}$ slower than we would expect given its amount of HI. There is some ambiguity to this statement due to the uncertainty in inclination for such a disturbed galaxy. However, even a broad range in inclinations will result in the same conclusion. In subsection \ref{sec:hicontent}, we made the opposite conclusion: namely that the NGC 3664 system had less HI than expected (see Figure \ref{fig:Pred_Mass}). This was based on the optical extents of the dwarfs in NGC 3664. The fact that the kinematic analysis above tells a different story, indicates that NGC 3664 is a perturbed system.

\quad In the left panel of Figure \ref{fig:pv_slice}, we show the position-velocity diagram for NGC 3664A from our VLA data. The diagram clearly illustrates that the rotation curve of the HI gas does not flatten out, but instead keeps rising. This is in contrast to the findings of \citet{swaters09}, who showed that in isolated late-type dwarf galaxies the flat part of the rotation curve was reached past 2 scale lengths of the optical disk. Interestingly, the HI component of NGC 3664A spans approximately 4 optical disk scale lengths. Additionally, the HI velocity field for NGC 3664A shows a very clear gradient across the galaxy, but it is not that of a normally differentially rotating disk, characterized by the two halves of the galaxy, as split by the kinematic minor axis, moving at plus/minus the approximate same velocity. This fact is also in contrast to the LMC, a similar galaxy in the optical, in which the flat part of the rotation curve is reached \citep{kim98}.

\quad NGC 3664A therefore presents two unique problems: 1) the dwarf has too much HI, when compared to the expected relations from \citet{karachentsev17}, 2) the dwarf does not flatten out kinematically as expected from \citet{swaters09}. One mechanism that could be causing these kinematic oddities, is an inclined warp, similar to that seen in M33, though at a different viewing angle \citep{corbelli97}. This is a hard effect to disentangle without rigorous modeling, as the optical morphology of the galaxy itself is quite complicated.

\quad Another possible explanation, is that the disturbed kinematics could be due to 
peculiar motions of gas within the galaxy itself. The Tully-Fisher relationship, used in
the \citet{karachentsev17} relation, may not be well calibrated for galaxies like NGC 
3664A, because the W$_{50}^{i}$ might not truly represent the kinematic spread within the galaxy. Sufficient amounts of gas could have been moved to the outskirts, which could add to the total mass but go unnoticed by a crude measure such as the W$_{50}^{i}$. The fact that the rotation curve does not flatten out, could also be due to the fact that there are flows of gas within the galaxy itself due to the ongoing interaction. This work thus shows the importance of fully understanding the environment of a galaxy and how the environment can challenge scaling relation expectations.

\quad For the NGC 3264 system (see Fig. \ref{fig:pv_slice}, middle), the rotation curve of the approaching side of the primary dwarf (left) appears to flatten out. The location of the secondary (see white circle), makes it difficult to determine whether the receding side also flattens out. Assuming an inclination of 90$^{\circ}$\footnote[2]{Measurement from the entry for NGC 3264 in Hyperleda \citep{hyperleda}.} for the primary galaxy, we calculate a W$_{50}^{i} = 133.9$ km s$^{-1}$, measured over the entire NGC 3264 system. This matches the expected width given the Tully-fisher relationship in \citet{karachentsev17} figure 9 to within less than 0.5$\sigma$, and its velocity map in \ref{fig:HI_Maps} looks typical for a differentially rotating system. As a result of these more regular kinematics, the dynamical state of NGC 3264 appears to be more relaxed than that of NGC 3664. This could be due to a number of reasons ranging from dynamical stage in the system's evolution, or, more likely, the fact that the NGC 3664 and NGC 3264 systems differ both in mass ratios of the member dwarfs and total distance separating them. Additionally, NGC 3264 has more regular contours, as opposed to the more more diffuse irregular contours at the lowest level for UGC 4638 and NGC 3664. This further indicates that the dynamical state of NGC 3264 is more relaxed than NGC 3664 and UGC 4638. 

\subsection{The Kinematics and Environment of the UGC 4638 Dwarf Triplet}
\label{sec:triplet}

\quad The UGC 4638 dwarf triplet is within 180 kpc (projected) from the massive galaxy UGC4640. This projected distance is likely to be within the virial radius of UGC4640\footnote{UGC4640 and the Milky Way both have R-band absolute magnitudes of approximately -21 mag. We therefore assume a Milky Way virial radius of approximately 200 kpc for this statement \citep{dehnen06}.}, and this together with the similar systemic velocities of the two suggest they may be interacting. The systemic velocities for UGC 4638 and UGC 4640 are 3315 $\pm$ 153 km s$^{-1}$ and 3312 $\pm$ 13 km s$^{-1}$, respectively. This indicates that the majority of the triplet's orbital motion around the more massive UGC 4640 is taking place in the plane of the sky or that we are catching the triplet close to its orbital apocenter. The HI enveloping the UGC 4638 system is compressed in the direction parallel to UGC 4640, and with a normal gradient, if not slightly extended, in the direction perpendicular to UGC 4640. This HI morphology additionally supports the claim of a plane of the sky orbit, and may be the result of tidal forces. Future dynamical modeling of the HI data can be used to disentangle the mutual interactions of the triplet and the massive galaxy. Such modeling can be compared to the results in \citet{besla12} for the SMC/LMC/MW system, as well as NGC 4490/85 in \citet{pearson18}. In Figure \ref{fig:pv_slice}, we also show the position-velocity diagram along the major axis of the HI morphological map for the UGC 4638 triplet (right panel), however little information can be taken from this diagram as the galaxies within the system are too close to distinguish individual kinematic features.

\quad The HI morphology of UGC 4638 suggests that the system is not being ram pressure stripped by the CGM of UGC 4640. If UGC 4638 were orbiting UGC 4640 in the plane of the sky, any effect due to ram pressure stripping would cause HI irregularities in a one sided extension emanating from the pair \citep[e.g.,][]{chung07}. If UGC 4638 was moving directly toward or away from us (at apocenter), we would see more gas spurious in velocity as it were being stripped. The lack of ram pressure stripping signatures may be due to the relatively large separation between UGC 4638 and UGC 4640, as they are currently separated by at least three times the separation of the MC/MW system. If the triplet moves closer to UGC 4640, it may be more easily ram-pressure stripped by the CGM. In Figure \ref{fig:HI_Maps}, we showed that the mutual tidal interactions between the dwarfs in the triplet has moved HI to the outskirts of the system. Thus, the future scenario of ram-pressure stripping of the triplet by the CGM of UGC 4640 is similar to the scenario in the simulations in \citet{marasco16}, where dwarf systems are more easily ram pressure stripped after undergoing a mutual tidal interaction prior to infall to a massive galaxy.

\quad Another interesting feature of the UGC 4638/40 system is that the HI mass measured by the GBT is approximately a factor two higher than the mass found in the VLA data. This discrepancy is not from calibration errors, since this effect would be at most at the level 5\% and would not cause a factor 2 difference. It is also unlikely to be due to additional dwarf galaxies within the GBT beam, as at this mass difference they would have been above the detection threshold of the VLA data. The HI mass difference indicates there is likely a diffuse HI reservoir, which the VLA did not detect do to the nature of interferometric response. This reservoir is likely extensions surrounding the UGC 4638 system, the UGC 4640 spiral, or both. A shared HI envelope could be evidence of the beginning of an interaction between UGC 4638 and UGC 4640. Future VLA-D array observations could help to further elucidate the global profile of this system. Additionally, observations with the Five hundred meter Aperture Spherical Telescope (FAST) could be very useful as they will have the same resolution as VLA-D array, while also detecting any extended emission that the VLA would miss.

\quad To date, there have been very few dwarf systems that have been observed with the level kinematic and morphological detail presented in this paper. As we obtain more observations of dwarf triplets, they can be used to test theories about hierarchical merging and give insight to the nature of hierarchical merging within dwarf groups \citep[e.g.,][]{stierwalt17}. The observations presented in this work are essential for accurate modelling (e.g., \citealt[][]{privon13,pearson18}) as they contain resolved kinematic information as well as coincident stellar and gaseous tidal features. Future models can be used to understand the baryon cycles in dwarf galaxy interactions as well as the dwarfs' ultimate dynamical fate.

\section{Conclusions}\label{sec:conc}

\quad In this paper, we have presented new VLA-C and D array observations of neutral hydrogen in three candidate dwarf interactions: two pairs and one triple, as well as in the spiral galaxy, UGC 4640, which is a likely host to the dwarf triplet. Our work presents the first resolved HI map of a dwarf triple to date, and our observations have higher sensitivity and resolution than previous studies. This provides a deeper look into the gas distribution and dynamics of dwarf systems and extends the sample of resolved interacting dwarf galaxies described in \citetalias{pearson16}. Our results can be summarized as follows:

\begin{enumerate}
  \item We detect HI bridges and envelopes connecting the dwarf galaxies within the NGC 3264, NGC 3664, and UGC 4638 systems. This provides clear evidence that the individual dwarfs are tidally interacting, and yields a nearby glimpse into a process that is cosmologically rare at present day.
  
  \item We overlay the HI morphologies on Pan-STARRS images, and detect diffuse, starlight extended from each of the dwarfs in the UGC 4638 triplet which follows the overall color distribution of the galaxies. In the east of the UGC 4638 triplet, this diffuse starlight is coincident with extended HI. The fact that both the stars and gas have been tidally extended to similar radii, despite the gas typically being more extended in dwarfs, could indicate that the smaller galaxy in the triplet has passed through the primary. In the Pan-STARRS and GALEX data, we additionally detect an unwinding spiral arm in NGC 3264. The GALEX UV images also show active star-formation within the HI bridges of two of the three systems.
  
  \item We compare the empirically expected HI mass to the measured HI mass for our three systems as well as for the \citetalias{pearson16} sample, and find that the single dish HI mass estimates for the dwarfs are comparable to those of isolated dwarfs \citep[][]{swaters02}. The three dwarf systems and the \citetalias{pearson16} dwarf pairs are not HI deficient, but a large fraction of the HI content resides in diffuse envelopes. Thus, we find that dwarf interactions do not unbind gas, but they park gas a large distance, which is consistent with \citet[][]{pearson16,pearson18}.
  
  \item We show that the expected results of standard scaling relationships (i.e. Tully-Fisher and the HI size-mass relationship) can differ from the measured values by an order of magnitude for dwarf pairs vs isolated dwarfs.

  \item The three dwarf systems all span a range of mass ratios and degrees of isolation. Yet, their kinematics all differ from what would be expected from non-interacting dwarf irregular galaxies. The coincident HI and stellar morphologies presented in this work, as well as the detailed resolved kinematic maps, provide exciting prospects for constraining future dynamical models of these systems.
  
\end{enumerate}

\quad We conclude that, in order to fully understand interacting dwarf systems, it is critical to have observations that are: 1) sensitive to large-scale diffuse emission 2) have the resolution to identify individual features, and 3) the sensitivity to detect neutral gas down to the 10$^{18-19}$ atoms cm$^{-2}$ level. This work highlights the importance of interferometric maps of suitable resolution and sensitivity to understand the baryon cycle in dwarf-dwarf interactions. 

\begin{acknowledgements}

\quad This work was partially supported by NSF grant AST-1715944. Support for this work was provided by NASA through the NASA Hubble Fellowship grant \#HST-HF2-51466.001-A awarded by the Space Telescope Science Institute, which is operated by the Association of Universities for Research in Astronomy, Incorporated, under NASA contract NAS5-26555. The Flatiron Institute is supported by the Simons Foundation. 

\end{acknowledgements}

\facilities{Karl G. Jansky Very Large Array \citep{evla}}
\software{Astropy \citep{astropy13}, CASA \citep{casa}}

\bibliography{references}{}
\bibliographystyle{aasjournal}

\end{document}